\documentclass[11pt,twocolumn]{article}

\usepackage[normalem]{ulem}
\usepackage{cancel}
\usepackage{soul}
\usepackage{dsfont}
\usepackage{mathrsfs}
\usepackage{natbib}
\usepackage{graphics}
\usepackage[T1]{fontenc}
\usepackage[latin1]{inputenc}
\usepackage{supertabular}
\usepackage{indentfirst} 
\usepackage{color}
\usepackage{oldgerm}
\usepackage{textfit}
\usepackage{eso-pic,graphicx}
\usepackage[amssymb]{SIunits} 
\usepackage{nicefrac} 
\usepackage{epsfig} 
\usepackage{fixfoot} 
\usepackage{footnote}
\usepackage{subfigure} 
\usepackage{esdiff}
\usepackage{latexsym,amsfonts,amsmath,amssymb,mathrsfs,bbold,mathtools,esint}
\usepackage{nicefrac} 
\usepackage{booktabs} 
\usepackage{array} 
\usepackage{paralist} 
\usepackage{verbatim} 
\usepackage{subfigure} 

\usepackage[ plainpages=false, bookmarks, bookmarksnumbered,
colorlinks, linktocpage=true, linkcolor=blue, citecolor=blue, filecolor=black,
urlcolor=blue]{hyperref}
\usepackage{memhfixc} 
\usepackage[font=small,labelfont=bf,labelsep=period]{caption}
\usepackage[affil-it]{authblk}
\usepackage[affil-it]{authblk}
\usepackage{blindtext}
\usepackage{abstract}

\usepackage{graphicx}
\usepackage{makeidx}
\usepackage{fancyhdr}
\usepackage{slashed}
\def\lambdabar{\protect\@lambdabar}
\def\@lambdabar{%
\relax
\bgroup
\def\@tempa{\hbox{\raise.73\ht0
\hbox to0pt{\kern.25\wd0\vrule width.5\wd0
height.1pt depth.1pt\hss}\box0}}%
\mathchoice{\setbox0\hbox{$\displaystyle\lambda$}\@tempa}%
{\setbox0\hbox{$\textstyle\lambda$}\@tempa}%
{\setbox0\hbox{$\scriptstyle\lambda$}\@tempa}%
{\setbox0\hbox{$\scriptscriptstyle\lambda$}\@tempa}%
\egroup
}

\usepackage[top=0.8in, bottom=0.7in, left=0.8in, right=0.8in]{geometry}

\begin{document} 
	 
\title{\bf Acceleration of electrons by high intensity laser radiation in a magnetic field}

\author{ {\bf Robert Melikian} \thanks{\rm E-mail address: robertmelikian@gmail.com}\\

\sl{A.I. Alikhanyan National Science Laboratory, Alikhanyan Brothers St. 2}\\
\sl{Yerevan 375036, Armenia}\\}
 
\date{\today} 

\twocolumn[
  \begin{@twocolumnfalse}
  \maketitle
\begin{abstract}
	We consider the acceleration of electrons in vacuum by means of the circularly-polirized electromagnetic wave, propagating along a magnetic field. We show that the electron energy growth, when using ultra-short and ultra-intense laser pulses ($1\:ps$, $10^{18}\:W/cm^2$, $CO_{2}$ laser) in the presence of a magnetic field, may reach up to the value $2,1\:Gev$. The growth of the electron energy is shown to increase proportionally with the increase of the laser intensity  and the initial energy of the electron. We find that for some direction of polarization  of the wave, the acceleration of electrons does not depend on the initial phase of the electromagnetic wave. We estimate the laser intensity, necessary for the electron acceleration. In addition, we find the formation length of photon absorption by electrons, due to which one may choose the required region of the interaction of the electrons with the electromagnetic wave and magnetic field. We also show that as a result of acceleration of electrons in the vacuum by laser radiation in a magnetic field one may obtain electron beam with small energy spread of the order $\nicefrac{\delta\epsilon}{\epsilon}\le 10^{-2}$.
\end{abstract}
\noindent \textbf{Keywords:} {Vacuum Electron Acceleration, High Intensity Laser.\\ \bigskip
}
\end{@twocolumnfalse}
]

\saythanks

\section{Introduction}\label{intro}
There has been a renewed interest to the topic of high intensity laser-accelerated electron beam in a magnetic field,  after the high intensity lasers, exceeding the values of the order $10^{18}\:W/cm^{2}$, have appeared. The latter open the possibility for multi-photon absorption and high-rate acceleration of electrons regime (Kolomenski \& Lebedev, \citeyear{Kolomenskii:1962ff}; \citeyear{Kolomenskii:1963ds}; \citeyear{Kolomenskii:1966gg};  Woodworth {\em et~al.}, \citeyear{Woodworth:1996ai}; Milantiev, \citeyear{Milantiev:1997jf}; Wang \& Ho \citeyear{Wang:1999wa}; Hirshfield \& Wang, \citeyear{Hirshfield:2000yz}; Milantiev \& Shaar, \citeyear{Milantiev:2000vc}; Konoplev \& Meyerhofer, \citeyear{Konoplev:2000uh}; Marshall {\em et~al.}, \citeyear{Marshall:2001zka}; Stupakov \& Zolotorev \citeyear{Stupakov:2000eu}; Singh, \citeyear{Singh:2004zz}; \citeyear{Singh:2006vv}; Cline {\em et~al.}, \citeyear{Cline:2013dd};  Galow {\em et~al.}, \citeyear{Galow:2013ff}). Due to their compactness such accelerators may be used in various applied areas, where the required energy of the electron beam is relatively small (of the order $100\:MeV$ to $1\:Gev$). The laser acceleration of electrons in electromagnetic field, in the presence of a magnetic field has been discussed in numerous papers by means of numerical analyses of the classical equations of motion of the electron. In particular, the electron acceleration under the influence of the intense linearly polarized laser beam in the presence of a magnetic field in vacuum due to the ponderomotive forces has been considered in (Woodworth {\em et~al.}, \citeyear{Woodworth:1996ai}). The dynamics of the electrons in the case of the focused laser beam  and plane wave was considered on the basis of the numerical calculations for the classical Hamilton equations. It was also shown there, that for the case of a focused laser beam the  maximal energy gain of electrons can be reached on the Rayleigh length.

During the process of electron acceleration in vacuum by means of the high intensity laser pulses and in the presence of a magnetic field, the interaction strength between the electron with the external fields may become so strong, that to account for the precise contributions of the resulting effects, it is necessary to consider the exact solution of the Dirac equation in these external fields (Redmond, \citeyear{Redmond:1965la}; Oleinik \& Sinyak, \citeyear{Oleinik:1975ba}; Abakarov \& Oleinik, \citeyear{Abakarov:1975nf}; Barber \& Melikian, \citeyear{Melikian:1999dy}). In this paper we estimate the energy growth of the electrons in external fields, based on the analytical investigation of the quasienergy spectrum  of electrons obtained in Barber and Melikian (\citeyear{Melikian:1999dy}), utilizing the  exact solution of the Dirac equation in these fields for an arbitrary laser intensity.

Our paper is organized as follows. In sections \ref{sec:quasi} and \ref{sec:growth} we show that during the interaction of an electron with a high intensity electromagnetic wave in a magnetic field, a substantial energy growth may happen due to the correction in the quasienergy spectrum of the electron, proportional to the laser intensity. As was shown in Barber and Melikian (\citeyear{Melikian:1999dy}), the dependence of the quasienergy spectrum of an electron on the polarization of the electromagnetic wave is crucial. In particular, for some particular direction of the polarization of an electromagnetic wave, the acceleration of electrons in vacuum by means of high intensity laser pulses in a magnetic field, is not sensitive to the initial phase of the electromagnetic wave. This has an essential practical advantage for implementing the process of acceleration. 

We also show in sections \ref{sec:quasi} and \ref{sec:growth} that the non-linear in  the wave field term in the expression for the quasienergy,  responsible for high rate of electron acceleration for intense laser pulses,  has a simple physical meaning. An electron interaction with the electromagnetic field in the presence  of a magnetic field is equivalent to the scattering of the electromagnetic wave on electrons, bound by the magnetic field. Within the quantum theory, electrons energy growth is the result of the momentum transfer from photons to electrons during the process of resonance photon absorption by electrons. It is clear that the probability of the resonance transitions of electrons (and, consequently, the energy growth) on the photon absorption length formation is many orders greater compared to the scattering of the electromagnetic wave on free electrons. Moreover, in section \ref{sec:growth} we consider the conditions necessary for the process of multi-photon  absorption by electrons.


 To estimate the necessary region of the electron interaction with the electromagnetic wave and the magnetic field, we also find in section \ref{sec:growth} the formation length corresponding to photon absorption by electrons. We also estimate the laser intensity, necessary for electron acceleration. These factors are essential for the realization of the electron acceleration by means of the laser pulses in a magnetic field.

 In section \ref{sec:rel} we show that due to electron acceleration in vacuum by means of high intensity laser pulses in a magnetic field, one may generate an electron beam with small energy dispersion, of the order $\nicefrac{\delta \epsilon}{\epsilon} \le 10^{-2}$. Finally, in conclusion, we give an overview of the obtained results.
 
\section{Quasienergy electron spectrum. Condition of radiation absorption by an electron.} \label{sec:quasi}

 The relativistic quantum states, as well as the quasienergy spectrum of an electron in the field of the circularly polarized wave of an arbitrary intensity, propagating along a homogeneous magnetic field, based on the exact Dirac solutions in these fields, have been considered in (Redmond, \citeyear{Redmond:1965la}; Oleinik \& Sinyak, \citeyear{Oleinik:1975ba}; Abakarov \& Oleinik, \citeyear{Abakarov:1975nf}; Barber \& Melikian, \citeyear{Melikian:1999dy}). This allows to reveal the specifics and various aspects of this interaction. In particular, the quasienergy spectrum, which was found in Barber and Melikian (\citeyear{Melikian:1999dy}) (without taking into account the anomalous magnetic moment of the electron), is given by the following expression:
\begin{align}
	Q^{\mu}_{q,n,\zeta}&=q^{\mu}+\frac{k^{\mu}}{(kQ)}|e|B(2n+1+\zeta) \notag \\ &+ k^{\mu}\frac{m^{2}\xi^{2}}{2\left((kQ)-g|e|B\right)}, \label{quasi:qenergy}
\end{align}
We use here the following notations: $Q^{\mu}_{q,n,\zeta}$ is the 4-quasimomentum of the electron, i.e., the time average of the kinetic 4-momentum of the electron in the state described by the quantum numbers $n,\zeta$, where $n=0,1,2,...$ is the level number of the discrete spectrum of the electron, in the direction perpendicular to the magnetic field $\vec{B}$, and $\zeta=\pm 1$ is the projection of the electron spin onto the direction of $\vec{B}$. The components $Q^{0}_{q,n,\zeta}$ and $Q^{3}_{q,n,\zeta}$ are the quasienergy and quasimomentum correspondingly. We have also denoted by $k^{\mu}(\omega,\omega\vec{n})$ the 4-vector of the electromagnetic wave, the frequency of the latter is denoted by $\omega$, and the unit vector in the direction of propagation of the wave is denoted by $\vec{n}$. The parameter $\xi=\nicefrac{eE}{m\omega}$ is the parameter of the wave intensity, and $E$ is the amplitude of the electric field of the electromagnetic field. Finally, $q^{\mu}$ is the 4-momentum of the free electron, which corresponds to the electron 4-momentum before entering to the region of interaction with the external fields. The values of $g=\pm 1$ determine the directions of the right or left circular polarization of the wave.\footnote{Everywhere, except when written explicitly, we use the units where $\hbar=c=1$.}

Some comments are in order. The second term in \eqref{quasi:qenergy} describes the interaction of an electron with the magnetic field. If one considers the limit $\xi \to 0$, then, taking into account the relations $q^2 =m^2$ and $k^2 =0$, one obtains from \eqref{quasi:qenergy} the well-known result for the electron energy in a constant magnetic field (Berestetskii {\em et~al.}, \citeyear{Berestetskii:1982qu}; Sokolov \& Ternov, \citeyear{Sokolov:1986ra}):
\begin{align}
	(Q^{0}_{q,n,\zeta})^{2}=(Q^{3}_{q,n,\zeta})^{2} +m^{2} +m\omega_{c}(2n+1+\zeta), \label{quasi:Q0Q3}
\end{align}
where $\omega_{c}=\nicefrac{eB}{m}$.

The third therm in \eqref{quasi:qenergy} describes the interaction of an electron with the electromagnetic wave and the magnetic field simultaneously. In particular, taking the limit $B \to 0$ one obtains from \eqref{quasi:qenergy} the well-known expression for the 4-quasimomentum of the electron in the electromagnetic field 
(Berestetskii {\em et~al.}, \citeyear{Berestetskii:1982qu}):
\begin{align}
	Q^{\mu} = q^{\mu} + k^{\mu}\frac{m^{2}\xi^{2}}{2(kq)}. \label{quasi:Qmu}
\end{align}

Using the condition $k^{2}=0$ one obtains from \eqref{quasi:qenergy} the relation $(kQ)=(kq)$ and the following expression for the quasienergy of the electron, in the state described by the quantum numbers $n,\zeta$:
\begin{align}
	Q^{0}_{q,n,\zeta} = q^{0} +\gamma_{0}\omega_{c}(2n+1+\zeta) + \frac{m\gamma_{0}\xi^{2}}{\delta}, \label{quasi:Q0}
\end{align}
where have denoted for brevity $\delta=1 - g\frac{2\gamma_{0}\omega_{c}}{\omega}$, and $\gamma_{0}=\nicefrac{(q^{0} + q^{3})}{2m}$.

After injecting the electrons into a magnetic field, they will occupy some interval of the quasienergy levels $Q^{0}_{q,n,\zeta}$, according to the spread of the values corresponding to $q^{0}$ and the quantum number $n$ of the electrons. It is clear from \eqref{quasi:qenergy} that the spread on the quantum number $n$ can be determined according to the spread of the electron velocity $V_{e}$ on angles $\phi$, where $\phi$ is the angle between the electron velocity $\vec{V}_{e}$ and the $z$-axis. 

From the expression \eqref{quasi:Q0} it follows that the quasienergy of the electron in external fields  depends essentially on the polarization $g$ of the electromagnetic field. It is clear from \eqref{quasi:Q0} that the corrections to the quasienergy of the electron, which are proportional to $\xi^{2}$, will always be positive if the polarization $g=-1$. Thus, by choosing appropriately the direction of the polarized wave, the growth of the electron energy will be independent on the initial phase of the electromagnetic wave. This is an essential circumstance from the practical point of view, when implementing the acceleration process.

\section{Electron energy growth. Photon absorption length formation.}\label{sec:growth}

From the formulas \eqref{quasi:qenergy} and \eqref{quasi:Q0} one sees that during the interaction of the electron with the electromagnetic wave and the magnetic field the energy growth may happen as a result of two effects. First, it may happen due to consecutive multiple resonance absorption of photons from the wave, during the transition between the quasienergies $Q^{\mu}_{q,n,\zeta} \to Q^{\mu}_{q',n',\zeta}$. Secondly, it may happen due to absorption of the $s$-photons from the wave, as a result of the correction to the electron energy, proportional to $\xi^{2}$.

\begin{figure}[htbp]
	\centering
		\includegraphics[height=1.1in]{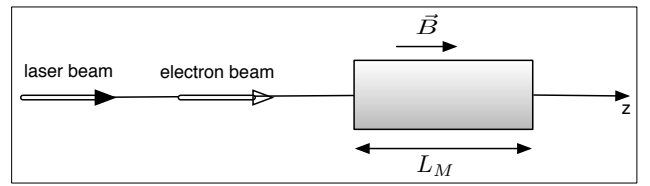}
	\caption{Schematic diagram of the electron acceleration by laser radiation in the presence of a magnetic field. $L_{M}$ is the length of the magnet.}
	\label{fig:laserbeam}
\end{figure}

The resonance absorption of a photon from the wave, with the 4-momentum $k^{\mu}(\omega,\vec{k})$, may occur during the transition between the quasienergies $Q^{\mu}_{q,n,\zeta} \to Q^{\mu}_{q',n',\zeta}$, if the length of the interaction (see Figure \ref{fig:laserbeam}) between the electron and the laser beam in the presence of a magnetic field is greater than the formation length $l_{a}$ of the absorption of a photon.
\newline
Due to the energy-momentum conservation, one obtains:
\begin{align}
	Q^{\mu}_{q,n,\zeta} + k^{\mu} = Q^{\mu}_{q',n',\zeta}.\label{growth:conservation}
\end{align}

The changes of the  quasienergy $\Delta Q^{0}$  and the quasimomentum $\Delta Q^{3}$ of an electron are related to the energy $\hbar \omega$ and the momentum $\hbar k_{z}$ of the absorbed photon. Using the conservation of the quasienergy-quasimomentum:
\begin{align}
	Q^{0}_{q,n,\zeta} + \omega &= Q^{0}_{q',n',\zeta} \label{growth:Q0} \\
	Q^{3}_{q,n,\zeta} + \omega \cos \theta &= Q^{3}_{q',n',\zeta}, \label{growth:Q3}
\end{align}
and the relation, following from \eqref{quasi:qenergy}:
\begin{align}
	\left(Q^{0}_{q,n,\zeta} \right)^{2} &= \left(Q^{3}_{q,n,\zeta} \right)^{2} +m^{2} +m \omega_{c}(2n+1+\zeta)\notag \\ &+\frac{m^{2}\xi^{2}}{\delta}, \label{growth:Q0Q3_relation}
\end{align}
we find the condition of the absorption of a photon by an electron:
\begin{align}
	Q^{0}_{q,n,\zeta} - (\cos \theta) Q^{3}_{q,n,\zeta} + \frac{\omega \sin^{2}\theta}{2} = \frac{m \omega_{c}(n'-n)}{\omega}. \label{growth:abs}
\end{align}
Here, $Q^{0}_{q,n,\zeta}$  and $Q^{3}_{q,n,\zeta}$ are the initial quasienergy and the $z$-component of the quasimomentum of the electron, and $\theta$ is the angle of the laser beam propagation, relative to the direction of the magnetic field $\vec{B}$.

We stress that we have considered here only the transitions between the quasienergies of the electrons without taking into account the changes in spin direction, since the probability associated with such changes is ignorable (Berestetskii {\em et~al.}, \citeyear{Berestetskii:1982qu}; Sokolov \& Ternov, \citeyear{Sokolov:1986ra}).

For the radiation absorption at optical or lower frequencies, and for the relevant to our case values of the angle $\theta \ll 1$, one can ignore in \eqref{growth:abs} the quantum corrections, since:
\begin{equation}
	\hbar \omega \sin^{2}(\theta) \ll \frac{mc^{2}\omega_{c}(n' - n)}{\omega}. \label{growth:limit}
\end{equation}
Thus, using the equation \eqref{quasi:Q0}, and substituting $Q^{3}_{q,n,\zeta}$ from \eqref{growth:Q0Q3_relation} into \eqref{growth:abs}, one can write the condition of the photon absorption in the form:
\begin{align}
	\omega \simeq \frac{\omega_{c}(n' - n)}{\gamma\left(1 - \cos(\theta)\sqrt{1- \frac{1}{\gamma^{2}} - \frac{ \omega_{c}(2n+1+\zeta)}{\gamma^{2}m} - \frac{\xi^{2}}{\gamma^{2}\delta}} \right)}, \label{growth:omega}
\end{align}
where we have introduced the notation:
\begin{align}
	\gamma= \frac{Q^{0}_{q,n,\zeta}}{m} =\gamma_{0}\left[ 1+ \frac{1}{4\gamma_{0}^{2}} + \frac{ \omega_{c}(2n+1+\zeta)}{m} + \frac{\xi^{2}}{\delta}\right]. \label{growth:gamma}
\end{align}
From the formula \eqref{growth:omega} one can easily obtain the classical condition  for the radiation absorption. Indeed, considering the transitions for the main harmonics corresponding to $n' - n =1$, disregarding the quantum correction $\nicefrac{2n\hbar \omega_{c}}{\gamma^{2}mc^{2}}$, and considering $\nicefrac{\xi^{2}}{\gamma^{2}\delta} \to 0$, one obtains the well-known classical condition for the radiation absorption (Kolomenskii \& Lebedev, \citeyear{Kolomenskii:1962ff}; \citeyear{Kolomenskii:1963ds}; \citeyear{Kolomenskii:1966gg}; Milantiev, \citeyear{Milantiev:1997jf}).

Since we are interested in the case for which  $\theta \ll 1$ and $\gamma_{0} \gg 1$, one can use the relations $(kQ) = (kq)$, as well as the formulas \eqref{growth:abs} and \eqref{growth:omega}, to find the approximate value of the difference $\Delta Q^{0} = Q^{0}_{q',n',\zeta} - Q^{0}_{q,n,\zeta}$ between the quasienergies of the electron:
\begin{align}
	\Delta Q^{0} \simeq \frac{2\gamma_{0}\omega_{c}(n' - n)}{1 + \theta^{2}\gamma_{0}^{2}\left[1 + \frac{\xi^{2}}{\delta} + \frac{ \omega_{c}(2n+1+\zeta)}{m} \right]}. \label{growth:Q0_approx}
\end{align}
It is clear from this formula, that the difference $\Delta Q^{0}$ between the quasienergies is not equidistant, and is decreasing with the increase of the quantum number $n$. Due to this behavior we see that for the quantum numbers $n > n_{cr}$, where $n_{cr}$ is some critical value, one will have $\Delta Q^{0} < \hbar \omega$. The latter means that the condition of the resonance absorption of a photon breaks down, and, as a result, the process of electron acceleration stops. We stress that $\Delta Q^{0}$, according to the formula \eqref{growth:Q0_approx}, does depend on the intensity parameter $\xi$, unless $\theta = 0$.

Let us also emphasize, that according to the formulas \eqref{growth:Q0} and \eqref{growth:Q3}:
\begin{align}
	Q^{0}_{q',n',\zeta} - Q^{0}_{q,n,\zeta} =\frac{Q^{3}_{q',n',\zeta} - Q^{3}_{q,n,\zeta}}{\cos(\theta)},\label{growth:Q0Q3_diff}
\end{align}
and, therefore, in the case when $\theta \ll 1$, the growth of the electron energy mainly occurs along the $z$-axis. This is in complete agreement with the fact that during the photon absorption the electron receives not only the energy, but also the momentum of the photon, along the direction of the $z$-axis.

From the condition for the photon absorption \eqref{growth:Q0_approx} it follows that besides the main harmonics $n' -n=1$, which corresponds to the classical physics, there are also possible the transitions with $n' -n \gg 1$. The probabilities of the quantum transitions with $n' -n \gg 1$ has been obtained and studied in (Abakarov \& Oleinik, \citeyear{Abakarov:1975nf}). It has been shown there, that if the quantum transition occur
under the influence of the pulsed electromagnetic field with the vector potential of the form $A=a exp(-\tau^2 /\tau_{0}^2)\cos(\omega t)$ (where $\tau_{0}$ is the duration of the wave pulse), then the probability of transitions $n \rightarrow n'$ is determined by the parameter:
\begin{equation}
\zeta_{tr} \simeq \sqrt{\pi} \xi \gamma_{0} \frac{\sqrt{\omega_{c}m}}{\omega} \tau_{0} \omega \exp \left[-\frac{\tau_{0}^2}{4}(2 \gamma_{0} \omega_{c} -\omega)^2 \right] \label{1:zeta}
\end{equation} 
In the paper (Abakarov \& Oleinik, \citeyear{Abakarov:1975nf}) it was shown that if $\zeta_{tr} \gg 1$ then the probability of transitions $n \rightarrow n'$ has the maximum value when $|n' -n| \sim \nicefrac{\zeta_{tr}^2}{2} \gg 1$. 
 
From the equation \eqref{growth:Q0_approx} it follows that if the energy of photons is equal to $\omega =2 \gamma_{0} \omega_{c}$, then during the trasition between the quasienergies $Q^{\mu}_{q,n,\zeta} \rightarrow Q^{\mu}_{q',n',\zeta'}$ there occurs $n' -n$-tuple photon absorption by electrons. Moreover, one can estimate the value of $|n' -n| \sim \nicefrac{\zeta_{tr}^2}{2}$ using the formula \eqref{1:zeta} for $\zeta_{tr}$, depending on the values of $\gamma_{0}$, $\xi$ and $\tau_{0}$. If $|1 - \nicefrac{2 \gamma_{0}\omega_{c}}{\omega}| \ll 1$, then one approximately obtains that $n' -n \sim \nicefrac{\zeta_{tr}^2}{2} \simeq \nicefrac{\omega}{(\pi |\omega -2 \gamma_{0}\omega_{c}|)}$ when the probability of transitions $Q^{\mu}_{q,n,\zeta} \rightarrow Q^{\mu}_{q',n',\zeta'}$ reaches maximum.

 In this paper we consider the accelration of electrons for  $g=-1$. However, according to the formula \eqref{quasi:Q0}, for the case  $g=+1$, the electrons energy growth is equal to $\nicefrac{m \gamma_{0} \xi^2}{(1 - \frac{2 \gamma_{0} \omega_{c}}{\omega})}$ and becomes stronger, in comparison with the case $g=-1$, at the resonance condition $\omega \rightarrow 2 \gamma_{0} \omega_{c}$. To understand and explain the peculiarities of the electron acceleration for the case $g=+1$ one needs to perform a more detailed analysis, which will be considered in a future publication.

As we have mentioned above, the absorption of a photon from the wave by an electron may occur only in the case when the interaction length of the electron with the laser beam in a magnetic field is greater the formation length $l_{a}$ of the photon absorption by an electron. To estimate the value of $l_{a}$ we note, that the process of photon absorption by an electron is formed on some definite length $l_{a}$ of the electron trajectory, when at least the wave with the length $\lambda$ may be absorbed during the time $\Delta t = \nicefrac{l_{a}}{V_{e}}$ (Landau \& Lifshitz, \citeyear{Landau:1975cl}; Berestetskii {\em et~al.}, \citeyear{Berestetskii:1982qu}; Sokolov \& Ternov, \citeyear{Sokolov:1986ra}). The length $l_{a}$ of the photon absorption is determined from the formula  (Landau \& Lifshitz, \citeyear{Landau:1975cl}; Sokolov \& Ternov, \citeyear{Sokolov:1986ra}):
\begin{align}
	l_{a} \simeq \frac{\lambda}{(1-\vec{\beta}\vec{n})},\label{growth:length}
\end{align}
where $\vec{\beta}=\nicefrac{\vec{V}_{e}}{c}$, and $\nicefrac{1}{(1-\vec{\beta}\vec{n})}$ is the Doppler factor. Clearly, in order the process of the photon absorption by an electron to be possible, the condition $l_{a}< L_{M}$, where $L_{M}$  is the length of the magnet, must be imposed. Representing $\vec{\beta}\vec{n}$ in the component form $\vec{\beta}\vec{n}=\beta_{z}n_{z}+\beta_{\perp}n_{\perp}=\beta_{z}\cos{\theta}+\beta\sin{\phi}\sin{\theta}$, and taking into account the formula \eqref{growth:length}, one obtains:
\begin{align}
	\beta_{z}\cos{\theta} \simeq 1-\frac{\lambda}{l_{a}}- \beta
\sin{\phi}\sin{\theta}. \label{growth:beta_z}
\end{align}
Using the relation $\gamma_{0}=\nicefrac{q^{0}(1+\beta_{z})}{2m}$, one can write the condition of the photon absorption \eqref{growth:Q0_approx}, for the case $\theta=0$, in the following form:
\begin{align}
	\omega = \frac{q^{0}}{m} \omega_{c}(1 + \beta_{z}).\label{growth:omega1}
\end{align}
Substituting $\beta_{z}$, for the case $\theta=0$, from the equation \eqref{growth:beta_z} into the above expression \eqref{growth:omega1}, we find:
\begin{align}
	l_{a} \simeq \lambda\frac{q^{0}\omega_{c}}{2q^{0}\omega_{c} - m\omega}.\label{growth:l_a_approx}
\end{align}
From the above expressions, using the limitation $0 < \beta_{z} <1$, one finds the following restriction on $\omega_{c}$:
\begin{align}
	\frac{m \omega}{2q^{0}} < \omega_{c} < \frac{m\omega}{q^{0}}.\label{growth:restriction_omega_c}
\end{align}
It is clear from \eqref{growth:l_a_approx} that for the given values of $\omega$ and $q^{0}$, by changing the values of $\omega_{c}$ within the limits determined in \eqref{growth:restriction_omega_c}, the length $l_{a}$ will be varying within the interval $(\lambda, \nu \lambda)$, where $\nu = \nicefrac{\tau_{0} c}{\lambda} = 1,2,...$ is determined from the duration $\tau_{0}$ of the laser pulse.

Let us note, that in the formulas  \eqref{growth:length}, \eqref{growth:beta_z} and \eqref{growth:l_a_approx} the  length $l_{a}$ represents the length of formation only of one photon. One can find the length $l_{a}$ of formation of photon absorption for $\xi \geq 1$, when the electron energy growth occurs because of absorption from the wave of $s$-photons, and the energy absorbed by the electron (on the wave length $\lambda$) is comparable or greater than the initial energy of the electron ((Landau \& Lifshitz, \citeyear{Landau:1975cl}). The overview of quantum processes in the field of a strong electromagnetic plane wave is given in (Nikishev \& Ritus \citeyear{Nikishov:1979ni}).

From \eqref{quasi:Q0}, \eqref{growth:Q0} and \eqref{growth:Q3} one finds that for case when $\xi \geq 1$, due to the interaction with the electromagnetic wave and magnetic field, the electron acquires the energy:
\begin{align}
	\Delta \epsilon =  Q^{0}_{q,n,\zeta} - q^{0} = s\omega \simeq \frac{m\gamma_{0}\xi^{2}}{1 + \frac{2 \gamma_{0}\omega_{c}}{\omega}}.\label{growth:delta_epsilon}
\end{align} 
From this formula it is clear that, for $\xi \geq 1$, the energy $\Delta \epsilon$ absorbed by an electron  may take values much greater than the initial electron energy $q^{0}$.

Using the relation \eqref{growth:Q0_approx} for the case $n'-n=1$ and $cos \theta=0$, valid for all values $\xi$, one obtains from \eqref{growth:delta_epsilon}:
\begin{align}
\frac{q^{0}(1+\beta_{z})\omega_{c}}{m\omega}=\frac{m\gamma_{0}\xi^{2}}{\Delta \epsilon} -1\label{growth:theta_0_cond}
\end{align}


Using the relation $\gamma_{0}=\nicefrac{q_{0}(1+\beta_{z})}{2m}$, and taking $\beta_{z} \simeq 1 - \frac{\lambda}{l_{a}}$, which follows from the formula \eqref{growth:beta_z} for the case $\theta = 0$, one obtains from \eqref{growth:delta_epsilon}:
\begin{align}
	\frac{2\Delta \epsilon}{\xi^{2}q^{0}} - 1 \simeq \beta_{z} \simeq 1 - \frac{\lambda}{l_{a}}.\label{growth:beta_z_approx}
\end{align}
From the formula \eqref{growth:beta_z_approx} we obtain the restrictions $\nicefrac{\xi^{2}q^{0}}{2} < \Delta \epsilon < \xi^{2}q^{0}$ and $l_{a} \geq l_{a,\text{min}} = \lambda$. In addition, we obtain the relation:
\begin{align}
	l_{a} \simeq \frac{\lambda}{2}\frac{\xi^{2}q^{0}}{\xi^{2}q^{0} -\Delta \epsilon}.\label{growth:l_a_2}
\end{align}	
The maximal value of the length $l_{a}$ is limited by the value of the laser pulse duration $\tau$, i.e., $l_{a,\text{max}} = \nu \lambda$. The dependence of $\Delta \epsilon$ on $l_{a}$, according to the formula \eqref{growth:beta_z_approx}, has the form:
\begin{align}
	\Delta \epsilon \simeq \xi^{2} q^{0}\left(1 - \frac{\lambda}{2 l_{a}} \right). \label{growth:delta_epsilon_l_a}
\end{align}

 In particular, in the case when the magnetic field $B \to 0$, one obtains from the formula \eqref{quasi:qenergy} the famous Dirac solution for an electron in the field of the plane electromagnetic wave (Berestetskii {\em et~al.}, \citeyear{Berestetskii:1982qu}). In this case, according to the expression \eqref{growth:delta_epsilon}, the energy growth  is equal to $m c^{2}\gamma_{0} \xi^{2}$. In the paper ( Stupakov \& Zolotorev, \citeyear{Stupakov:2000eu})  the authors had considered the acceleration of electrons interacting with a focused high intensity short duration laser beam in vacuum, by means of ponderomotive light forces. It was shown that the maximal energy gain of electrons can reach up to the value $m c^{2} \gamma_{0} \xi^{2}$ on the Rayleigh length of the order $\gamma_{0} \xi \lambdabar$.

To estimate the necessary laser intensity $I_{\text{las}}$ we use the relation between $I_{\text{las}}$ and the amplitude $E$ of the electromagnetic wave's electric field (Landau \& Lifshitz, \citeyear{Landau:1975cl}):
\begin{align}
	E \left[\frac{\text{V}}{\text{cm}}\right] \simeq 19,46 \sqrt{I_{\text{las}} \left[\frac{\text{W}}{\text{cm}^{2}}\right]}. \label{growth:laser_int}
\end{align}

\begin{figure}[h!]
	\centering
		\includegraphics[height=1.7in]{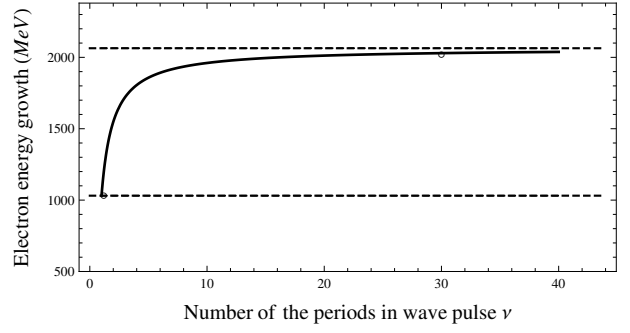}
	\caption{Dependence  of the electron energy growth $\Delta \epsilon$  on the number $\nu$ of periods in the wave pulse according to the formula \eqref{growth:delta_epsilon_l_a}, for the parameters $\gamma_{0}=10^{2}$, $I_{\text{las}} \simeq 10^{18}\:\nicefrac{\text{W}}{\text{cm}^{2}}$ (or $\xi \simeq 6.42$), $\lambda=1,06 \cdot 10^{-3}\:cm$.}
	\label{fig:delta_epsilon}
\end{figure}

Using \eqref{growth:delta_epsilon_l_a} and \eqref{growth:laser_int} we see that the energy growth $\Delta \epsilon$ increases proportionally with the increase of the laser intensity $I_{\text{las}}$ and the initial electron energy $q^{0}$. To give a concrete numeric estimate, we take, for example, $\gamma_{0} =10^{2}$, $\lambda=1,06 \cdot 10^{-3}\:cm$, $I_{\text{las}} \simeq 10^{18}\:\nicefrac{\text{W}}{\text{cm}^{2}}$, $\tau = 1\:ps$, we find that the energy growth has, according to the formula \eqref{growth:delta_epsilon_l_a}, the value $\Delta \epsilon \simeq 2,1\:GeV$ (see Figure  \ref{fig:delta_epsilon})

\section{Relative precision of the accelerated electrons energy.}\label{sec:rel}

For the case we consider in this paper, the parameters are $\xi \geq 1$, $g=-1$, and the electron energy growth is determined from the formula \eqref{growth:delta_epsilon_l_a}. The relative precision of an accelerated electron beam can be found from \eqref{growth:delta_epsilon_l_a}, using the following approximate formula:
\begin{align}
	\frac{\delta \epsilon}{\epsilon} \simeq 2 \frac{\delta \xi}{\xi} + \frac{\delta q^{0}}{q^{0}} +  \frac{1}{2\nu-1} \frac{\delta \tau_{0} }{\tau_{0}}.\label{rel:approx}
\end{align}
We see from this formula, that $\nicefrac{\delta \epsilon}{\epsilon}$ depends on the relative precision of the initial electron beam energy, the intensity, as well as the duration of the laser pulse. For example, taking $ \nicefrac{\delta q^{0}}{q^{0}} \leq 10^{-3}$, $\frac{\delta \omega}{\omega} \leq 10^{-3}$,  and $\nicefrac{\delta \xi}{\xi} \leq 0,5 \cdot 10^{-2}$,  one obtains the order of the relative precision of the accelerated electron beam: $\nicefrac{\delta \epsilon}{\epsilon} \le 10^{-2}$.

\section{Conclusion}

In this paper we have shown that for $I_{\text{las}} = 10^{18}\:\nicefrac{\text{W}}{\text{cm}^{2}}$, $\tau_{0} = 1\:ps$, $\gamma_{0}=10^{2}$, $\lambda = 1,06 \cdot 10^{-3}\:cm$,  the growth of the electron energy  is equal to $\Delta \epsilon \simeq 2,1\:GeV$, if one is to use the method of electron acceleration by laser radiation in a magnetic field.

We have also shown that the electron energy growth increases proportionally to the increase of the laser intensity and the initial energy of the electron. Moreover, we have  estimated the intensity of the laser $I_{\text{las}}$, necessary for the acceleration of electrons.  We find the formation length $l_{a}$ of photon absorption by electrons, on basis of which one may choose the necessary region of the interaction of the electrons with the electromagnetic wave and magnetic field.

We stress that to implement the acceleration process, it is crucial that for an appropriate choice of the wave polarization direction, namely, for $g=-1$, the acceleration of the electrons does not depend on the initial phase of the electromagnetic wave. In addition, as a result of the electron acceleration in vacuum by means of high intensity laser pulses in a magnetic field, one may obtain an electron beam with small energy spread, of the order $\nicefrac{\delta \epsilon}{\epsilon} \le 10^{-2}$.

\section*{Acknowledgments} The author would like to thank Prof. A.A. Chilingarian and Dr. S.P. Taroian for their support and interest to this work.

\end{document}